\def \SAIT #1 #2 {{\em Mem.\ Soc.\ Astron.\ It.\/} {\bf #1}, #2}
\def \MESS #1 #2 {{\em The Messenger\/} {\bf #1}, #2}
\def \ASTRNACH #1 #2 {{\em Astron. Nach.\/} {\bf #1}, #2}
\def \AAP #1 #2 {{\em Astron. Astrophys.\/} {\bf #1}, #2}
\def \AAL #1 #2 {{\em Astron. Astrophys. Lett.\/} {\bf #1}, L#2}
\def \AAR #1 #2 {{\em Astron. Astrophys. Rev.\/} {\bf #1}, #2}
\def \AAS #1 #2 {{\em Astron. Astrophys. Suppl. Ser.\/} {\bf #1}, #2}
\def \AJ #1 #2 {{\em Astron. J.\/} {\bf #1}, #2}
\def \ANNREV #1 #2 {{\em Ann. Rev. Astron. Astrophys.\/} {\bf #1}, #2}
\def \APJ #1 #2 {{\em Astrophys. J.\/} {\bf #1}, #2}
\def \APJL #1 #2 {{\em Astrophys. J. Lett.\/} {\bf #1}, L#2}
\def \APJS #1 #2 {{\em Astrophys. J. Suppl.\/} {\bf #1}, #2}
\def \APSS #1 #2 {{\em Astrophys. Space Sci.\/} {\bf #1}, #2}
\def \ASR #1 #2 {{\em Adv. Space Res.\/} {\bf #1}, #2}
\def \BAIC #1 #2 {{\em Bull. Astron. Inst. Czechosl.\/} {\bf #1}, #2}
\def \JSQRT #1 #2 {{\em J. Quant. Spectrosc. Radiat. Transfer\/} {\bf #1}, #2}
\def \MN #1 #2 {{\em Mon. Not. R. Astr. Soc.\/} {\bf #1}, #2}
\def \MEM #1 #2 {{\em Mem. R. Astr. Soc.\/} {\bf #1}, #2}
\def \PLR #1 #2 {{\em Phys. Lett. Rev.\/} {\bf #1}, #2}
\def \PASJ #1 #2 {{\em Publ. Astron. Soc. Japan\/} {\bf #1}, #2}
\def \PASP #1 #2 {{\em Publ. Astr. Soc. Pacific\/} {\bf #1}, #2}
\def \NAT #1 #2 {{\em Nature\/} {\bf #1}, #2}

\documentstyle{memsait}
\input epsf.sty
\begin{opening}
\title{RESOLUTION OF THE AGE DISCREPANCIES IN PULSAR/SNR ASSOCIATIONS}
\author{David Marsden$^1$, Richard E. Lingenfelter$^2$, Richard E. Rothschild$^2$}
\institute{$^1$NASA/Goddard Space Flight Center, Greenbelt, MD\\
$^2$UCSD/Center for Astrophysics and Space Sciences, La Jolla, CA}
\date{} 
\end{opening}

\begin{document}

\oddpagefooter{}{}{} 
\evenpagefooter{}{}{} 
\

\begin{abstract}
Pulsars associated with supernova remnants (SNRs) are valuable
because they provide constraints on the mechanism(s) of pulsar
spin-down. Here we discuss two SNR/pulsar associations in which
the SNR age is much greater than the age of the pulsar obtained by
assuming pure magnetic dipole radiation (MDR) spin-down. The PSR
B1757$-$24/SNR G5.4$-$1.2 association has a minimum age of $\sim$40 
kyr from proper motion upper limits, yet the MDR timing age of the
pulsar is only 16 kyr, and the newly discovered pulsar PSR J1846$-$0258 
in the $>$2 kyr old SNR Kes 75 has an MDR timing age of just 0.7 kyr. 
These and other pulsar/SNR age discrepancies imply that the pulsar 
spin-down torque is not due to pure MDR, and we discuss a model for 
the spin-down of the pulsars similar to the ones recently proposed 
to explain the spin-down of soft gamma--ray repeaters (SGRs) and 
anomalous x--ray pulsars (AXPs).
\end{abstract}

\section{Introduction}
The study of pulsars and their spin-down provides important
information on the physics of neutron stars. This information (e.g. 
magnetic field, moment of inertia, etc.) can be gleaned from the 
pulsar spin-down by assuming a physical model for the spin-down torque. 
For isolated pulsars, it is usually assumed that the spin-down torque 
is due to magnetic dipole radiation (MDR), which produces a timing 
age $\tau_{MDR}=0.5P/\dot{P}$ and pulsar braking index $n=\Omega
\ddot{\Omega}/\dot{\Omega}^2=3$, where $P=2\pi/\Omega$ and 
$\dot{P}=-2\pi \dot{\Omega}/\Omega^2$ are the pulsar spin
period and period derivative, respectively (this calculation of 
$\tau_{MDR}$ assumes that the initial spin period was much smaller 
than the observed spin period). Assuming MDR spin-down, the surface
magnetic field strength of the neutron star is given by the
formula $B=3.2\times10^{19}(P\dot{P})^{1/2}$G, which has been
widely used to estimate the magnetic field strengths of isolated
pulsars (e.g. Manchester \& Taylor 1977).

\section{Pulsar/SNR Age Discrepancies?}
Perhaps the only way to test the MDR spin-down hypothesis for
pulsars is to study the SNRs associated with young pulsars,
because the SNRs provide an age constraint independent of the
pulsar spin-down. There are at least two pulsars with SNR ages
which are inconsistent with the timing ages of the pulsars
calculated assuming MDR spin-down: PSR B1757$-$24 and PSR
J1846$-$0258. PSR B1757$-$24 is a 0.125 s radio pulsar associated with
SNR G5.4$-$1.2. Given the displacement of the pulsar from the center
of G5.4$-$1.2, the lack of observed proper motion of the pulsar 
with respect to its SNR implies an age $\tau_{psr}>$ 39 kyr 
(Gaensler \& Frail 2000), which is more than a factor of two 
greater than the MDR timing age of 16 kyr. Because the pulsar/SNR 
association seems so compelling, Gaensler \& Frail (2000) 
suggested that this age discrepancy raises very serious 
problems for all pulsar ages based on MDR.

Recently, the 0.324 s x-ray pulsar PSR J1846$-$0258 was discovered
(Gotthelf et al. 2000) in the center of the SNR Kes 75. The MDR
timing age of the pulsar is $\tau_{MDR}$= 0.7 kyr, which is
much less than the estimated minimum age of Kes 75. A lower limit 
to the age of the SNR can be estimated by assuming that the remnant 
is still undergoing free expansion. In this case,
\begin{equation}
M_{ej}(>v)>\frac{4}{3} \pi R^3 \rho_{ISM},
\end{equation}
where $R$ is the radius of Kes 75 (9.7 pc; Blanton \& Helfand
1996), $\rho_{ISM}$ is the density of the interstellar medium
surrounding Kes 75, and $M_{ej}(>v)$ is the total mass of ejecta
with free expansion velocities greater than $v$. Realistic models
of the velocity profiles of Type II supernova ejecta in free 
expansion consist of a broken power law distribution of ejecta 
velocities, such that the majority of the mass and energy is in 
the low-velocity ejecta. As a function of the ejecta power law 
index $q$ (index $n$ in Truelove \& McKee 1999),
\begin{equation}
\frac{M_{ej}(>v)}{M_{tot}} = 1 - \frac{(v/v_{core})^{3-q} -
q/3}{(v_{core}/v_{max})^{q-3} - q/3},
\end{equation}
where $M_{tot}$ is the total ejecta mass, $v_{max}$ is the maximum
ejecta velocity, and $v_{core}$ is the transition velocity below
which the distribution of ejecta velocities (in velocity space) is 
flat. The parameter $v_{core}$ can be eliminated by normalizing Equation 
(2) to a total ejecta kinetic energy $E_{tot}$, and Equations ($1-2$) 
can then be solved for the cutoff velocity $v$ for a given $\rho_{ISM}$, 
$R$, $q$, $M_{tot}$, and $E_{tot}$ (the result is insensitive to the value 
of $v_{max}$). The minimum age in free expansion is then given simply 
by $T>R/v$. For $q=9-10$ (Chevalier \& Fransson 1994), $E_{tot}=10^{51}$ 
ergs, $R$= 9.7 pc, $M_{tot}$= 20 $M_\odot$, $v_{max}$= 20,000 km s$^{-1}$, 
and assuming the SNR is in the hot, most tenuous phase of the ISM with 
$n_{ISM}$= 0.001, we estimate a minimum age $\tau_{snr}>$ 2.0 kyr for 
Kes 75. This is a factor of three greater than the pulsar's MDR timing 
age of 0.7 kyr.

What conditions must be met for Kes 75 to have the same age as the
MDR timing age of PSR J1846$-$0258? The minimum age estimate above
is insensitive to $v_{max}$, but decreases as $E_{tot}$ increases 
or $M_{tot}$ decreases. For $q=9-10$ and $n_{ISM}= 0.001$ cm$^{-3}$, 
and assuming a minimum Type II supernova ejecta mass of $\sim$6.6 M$_\odot$ 
(for an 8 M$_\odot$ progenitor star forming a 1.4 M$_\odot$ neutron star), 
we find that $\tau_{snr} = \tau_{MDR}$ for Kes 75 only if E$>4\times 
10^{51}$ ergs, which is roughly three times greater than the energy 
implied by model Type II supernova lightcurves (Shigeyama \& Nomoto 1990). 
Even for extremely high $E_{tot}$, $\tau_{snr} = \tau_{MDR}$ seems 
unlikely because the ISM density surrounding Kes 75 is probably much
greater than 0.001 cm$^{-3}$. This is because an OH maser emission
line has been observed from the SNR (Green et al. 1997), which is
thought to result from the supernova shock interacting with a
dense molecular cloud. Therefore, $\tau_{snr} >$ 2 kyr is probably
a very conservative lower limit to the Kes 75 age, since increasing 
$\rho_{ism}$ increases the minimum age from the arguments above. 

\section{Propeller-aided spin-down}
The age discrepancy problems outlined above suggest that simple
MDR spin-down is incorrect for these pulsars, and that exploration
(e.g. Marsen, Lingenfelter, \& Rothschild 2001) of more complete 
spin-down models is warranted. We consider a hybrid spin-down model 
consisting of MDR torques plus the addition of spin-down torque due 
to the ``propeller effect'' (Illarionov \& Sunyaev 1975) from material 
at the pulsar magnetosphere. In the case of PSRs B1757$-$24 and 
J1846$-$0258, this material could be supernova ejecta captured by 
the neutron star in the form of a {\it fallback disk}. Fallback disks 
may be roughly divided into two categories: ``prompt'' and ``delayed''. 
Prompt disks may be formed from $\sim 0.001-0.1M_{\odot}$ (Michel 1991; 
Lin, Woosley, \& Bodenheimer 1991) of ejecta material soon after the 
initial core collapse in a type II supernova explosion (Woosley \& 
Weaver 1995). Formation of such prompt disks is probably limited to 
$<7$ days after the core collapse because of heating of the ejecta 
by $^{56}$Ni decays (Chevalier 1989). Delayed disks may form years 
after the explosion from ejecta decelerated by a strong reverse shock 
(Truelove \& McKee 1999) caused by the primary supernova blast wave 
impinging on dense circumstellar material from the pre-supernova stellar 
wind. Such models were recently invoked to explain the spin-down of AXPs 
(Chatterjee, Hernquist \& Narayan 2000), and SGRs and AXPs (Marsden 
et al. 2001). 

\begin{figure}
\epsfysize=6cm 
\epsfbox{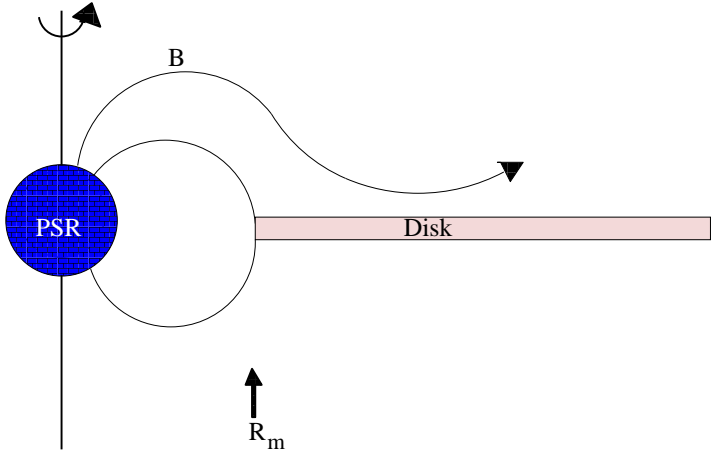}
\caption[h]{A pulsar with an accretion disk. The disk is 
truncated at an inner radius $R_{m}$.}
\end{figure}
An artist's conception of the pulsar and accretion disk system is 
shown in Figure 1. The total spin-down rate of the neutron star due to 
the combined disk and MDR torque is given by
\begin{equation}
\dot{\Omega} = \dot{\Omega}_{MDR} + \dot{\Omega}_A
\end{equation}
The MDR torque $I_{\ast}\dot{\Omega}_{MDR}=-B_{\ast}R^6_*\Omega^3/6c^3$ 
(e.g. Manchester \& Taylor 1977), where $I_{\ast}$ is the neutron star 
moment of inertia, $R_{\ast}$ is the neutron star radius, and the propeller
torque $I_{\ast}\dot{\Omega}_A=k\dot{m}R^2_m\Omega_{eq}(1-\Omega/\Omega_{eq}$)
(Menou et al. 1999), where $k$ is a constant of order unity, $\dot{m}=
10^{16}\dot{m}_{16}$ g s$^{-1}$ is the mass infall rate at the 
magnetosphere, and $R_m$ is the magnetospheric radius. Here and 
elsewhere we assume $R_{\ast} = 10$ km, $I_{\ast} = 1.1\times 10^{45}$ 
g cm$^2$, and $B_{\ast}=10^{12}B_{12}$ G is the surface magnetic field 
strength (assumed dipole). The equilibrium angular frequency $\Omega_{eq}$ 
is defined by the condition $R_m = R_c$, where $R_c = {(GM_{\ast}/
\Omega^{2})}^{1/3}$ is the Keplerian co-rotation radius for a neutron 
star of mass $M_{\ast}$. The timing age $\tau_{comb}$ under the action 
of the combined torque model is then given by
\begin{equation}
\tau_{comb} = \int_{\Omega_0}^{\Omega}
\frac{d\Omega}{\dot{\Omega}_{MDR} + \dot{\Omega}_A},
\end{equation}
where $\Omega_0 = 2\pi/P_0$ is the initial angular frequency. The timing 
ages were calculated for PSR B1757$-$24 ($P$= 0.125 s) and 
PSR J1846$-$0258 ($P$ = 0.324 s) for a grid of $B_{\ast}$ and $\dot{m}$ 
values using Equations ($3-4$) and assuming $R_M=2.4 \times 10^8 
B^{4/7}_{12} \dot{m}^{-2/7}_{16}$ cm (Chatterjee, Hernquist \& 
Narayan 2000).

\begin{figure}
\epsfysize=8cm 
\epsfbox{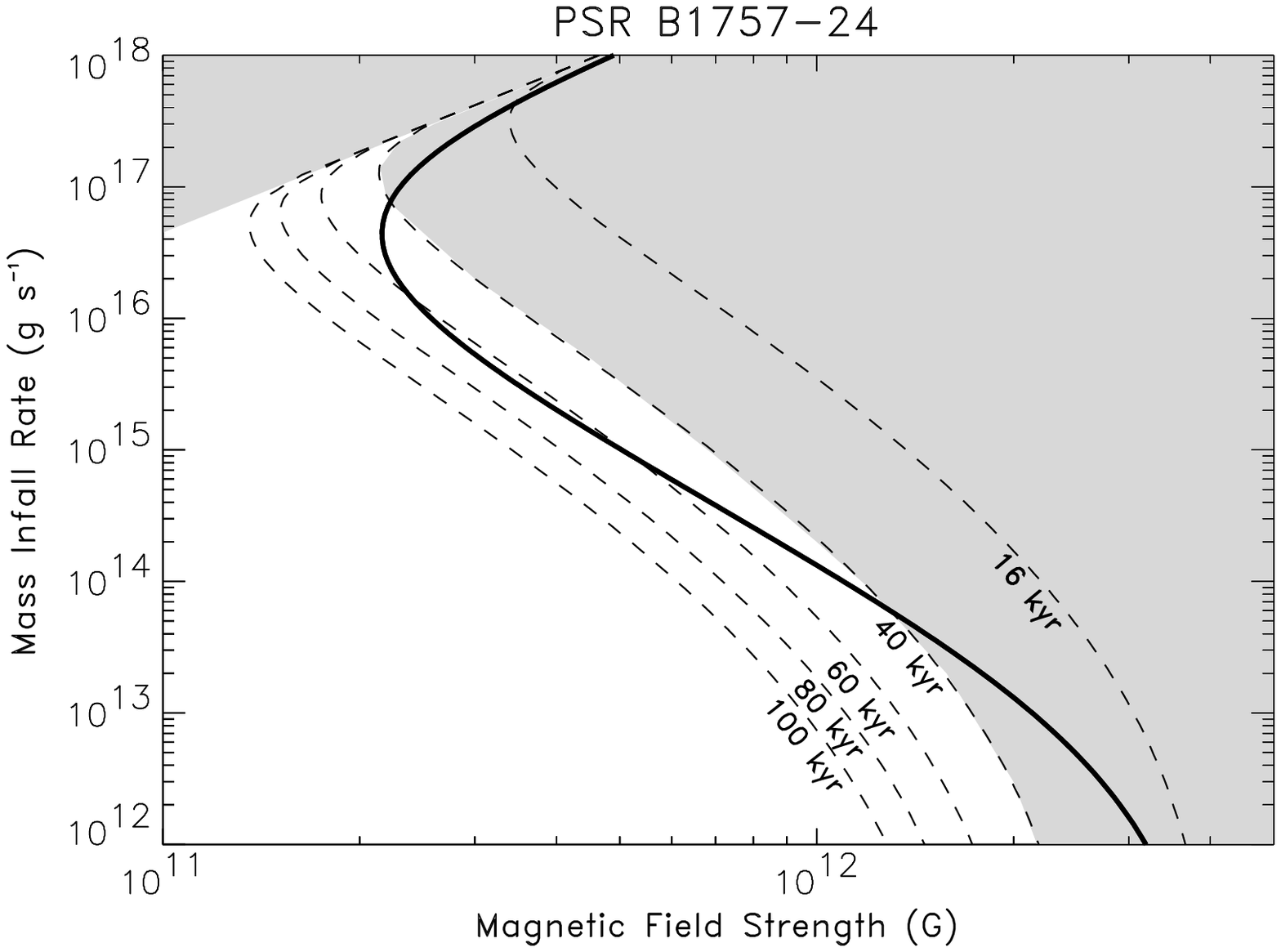}
\epsfysize=8cm
\epsfbox{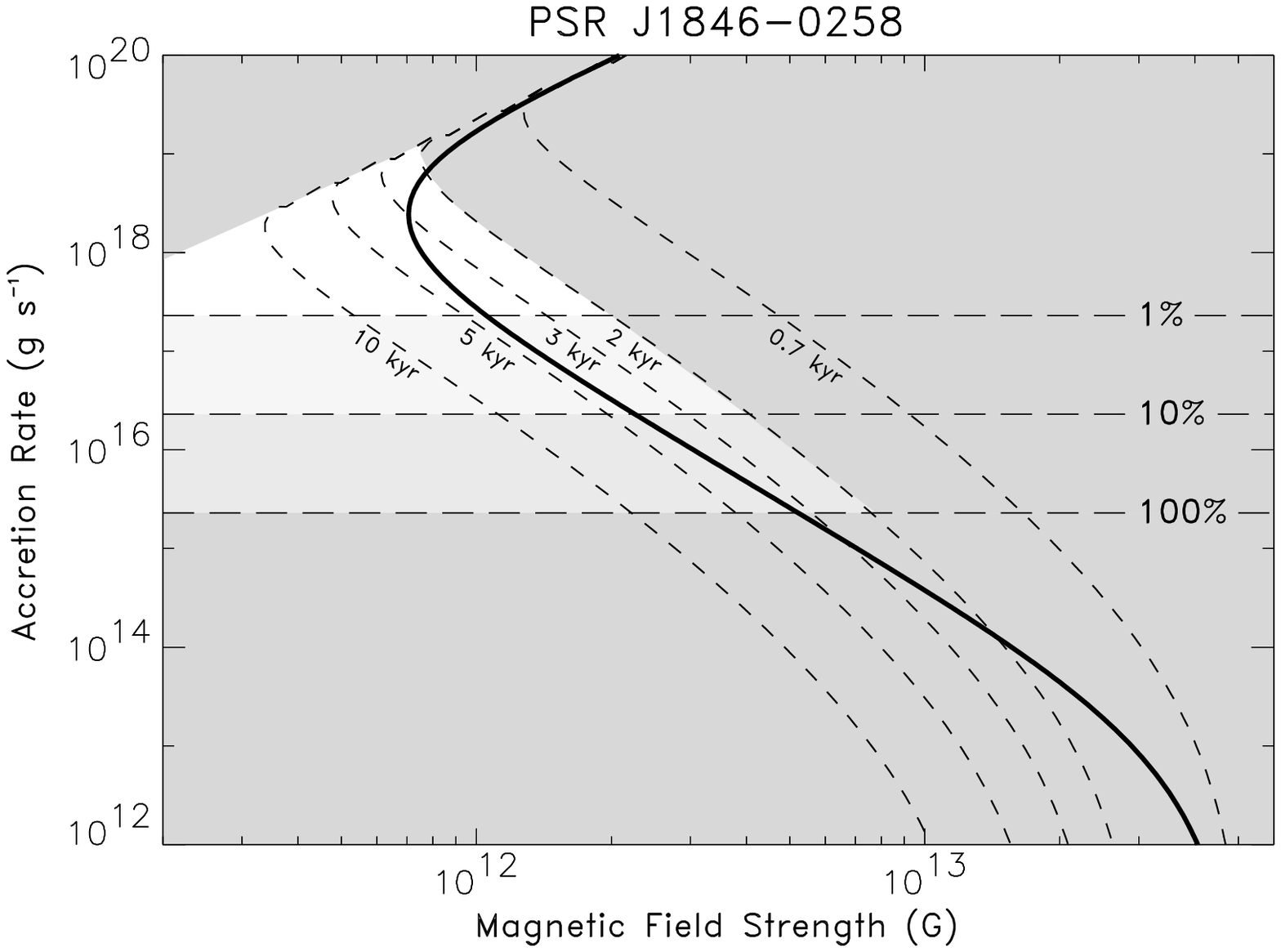}
\caption[h]{Contour plots of the PSR B1757$-$24 (top) and PSR J1846$-$0258 
(bottom) characteristic ages for the combined MDR and propeller torques 
spin-down model, for various values of the neutron star magnetic field 
$B_{\ast}$ and magnetospheric accretion rate $\dot{m}$. The allowed values 
of $B_{\ast}$ and $\dot{m}$ lie along the heavy solid lines corresponding 
to the present-day $\dot{P}$, and the shaded areas are excluded by the SNR 
age constraints or the condition $\dot{P} > 0$. For PSR J1846$-$0258, the 
dominant constraint on $B_{\ast}$ and $\dot{m}$ are provided by limits on 
the x-ray efficiency given the observed x-ray luminosity (see text).}
\end{figure}

\section{Results}
The resulting range of pulsar magnetic fields and magnetospheric
accretion rates are shown in Figure 2. For PSR B1757$-$24 (top
panel), we find that values of $9 \times 10^{10} < B_{\ast} < 1.4 \times 
10^{12}$ G, $7 \times 10^{13} < \dot{m} < 9 \times 10^{17}$ g s$^{-1}$, 
and $39 < \tau_{psr} < 80$ kyr are consistent with the lower limit on
the true age of 39 kyr (Gaensler \& Frail 2000) and the present
day spin-down rate of the pulsar (heavy solid line). Similarly,
for PSR J1846$-$0258 (right panel) the lower limit on the age ($\tau_{snr} 
>$ 2 kyr) gives a range of values $9 \times 10^{11} < B_{\ast} < 1.5 
\times 10^{13}$ G and $1.1 \times 10^{14} < \dot{m} < 8 \times 10 ^{18}$ 
g s$^{-1}$. In addition, a more stringent constraint is provided by
the x-ray luminosity of the pulsar, since (as seen from the
Figure) the magnetic field of the pulsar must be less than $\sim 1.5 
\times 10^{13}$G for consistency with the 2 kyr age lower limit. The MDR
luminosity for this magnetic field strength is only $L_{MDR} \sim 8 
\times 10^{35}$ ergs s$^{-1}$ --- much less than the observed x-ray 
luminosity of the pulsar ($L_x \sim 2 \times 10^{36}$ ergs s$^{-1}$; 
Gotthelf et al. 2000). Therefore the x-ray luminosity may be 
predominantly powered by the magnetospheric accretion in the context 
of this model, which constrains $\dot{m}$ for a given x-ray efficiency
defined by $L_x = \epsilon \dot{m} c^2$. For $\epsilon <$ 1,
Figure 2 (bottom) indicates that a solution exists for $7 \times 
10^{11}<B_{\ast}<6\times 10^{12}$G, $4\times 10^{15}<\dot{m}<8 
\times 10^{18}$ g s$^{-1}$, and $2< \tau_{snr}<5$ kyr. Since 
most of the accretion to the neutron star surface would be inhibited 
by the centrifugal barrier in these sources, the value of the x-ray 
efficiency is probably much less than the usually assumed value of 
$\epsilon = GM/Rc^2 = 0.2$ appropriate for x-ray binaries.

\section{Discussion}
The discrepancies between the MDR timing ages and supernova
remnant ages for pulsars B1757-24 and J1846-0258 can be resolved
by using a more complete spin-down model consisting of both MDR
and propeller torques. One prediction of this model is the
presence of excessive pulsar timing noise which is characteristic
of noisy propeller torques. In addition, optical or infrared
emission from isolated neutron star accretion disks may be
detectable (Perna, Hernquist \& Narayan 2000). Since polar cap
accretion is suppressed by the centrifugal barrier for $\Omega>
\Omega_{eq}$, pulsed radio emission from open field lines above
the disk is not precluded by this model (see e.g. Michel 1991, ch.
6). More work is needed, however, to incorporate the effects of 
{\it time-dependent} magnetospheric accretion, as the accretion 
rate should gradually decrease with time as the disk dissipates 
(Cannizzo, Lee \& Goodman 1990). Propeller-based spin-down models 
incorporating a time-dependent $\dot{m}$ have been proposed for 
AXPs (Chatterjee, Hernquist \& Narayan 2000), but this model did 
not include MDR spin-down torque, which may be significant for 
older pulsars (such as B1757-24) and pulsars with small initial 
$\dot{m}$. If a significant fraction of pulsars are born with 
accretion disks, then there may be a population of older pulsars 
--- whose disks have dissipated --- with abnormally {\it long} 
MDR timing ages. From just the simple model considered here, 
spin-down models incorporating propeller torques hold much 
promise for reconciling the supernova remnant ages and the 
timing parameters of pulsars.

\acknowledgements
This work was performed while one of the authors (DM) held a
National Research Council-GSFC Research Associateship. RER
acknowledges support by NASA contract NAS5-30720, and REL support
from the Astrophysical Theory Program.
%

\end{document}